\documentclass[pre,
twocolumn,
showpacs,amsmath,amssymb]{revtex4}
\usepackage{graphicx}
\usepackage{amsfonts}
\usepackage{times}
\usepackage{bm}
\usepackage{natbib}

\newcommand{\ba}[1]{\bm{a}^{(#1)}}

\newcommand{\bu}{\bm{u}}
\newcommand{\bx}{\bm{x}}
\newcommand{\bg}{\bm{g}}
\newcommand{\bH}[1]{{\cal H}^{(#1)}}
\newcommand{\bxi}{\bm{\xi}}

\newcommand{\pp}[2]{\frac{\partial #1}{\partial #2}}

\newcommand{\suma}{\sum_{a=1}^d}
\newcommand{\Kn}{\mbox{\sl Kn}}
\newcommand{\Ma}{\mbox{\sl Ma}}
\newcommand{\Hs}[1]{\mathbb{H}^#1}

\begin{document}

\title{Efficient kinetic method for fluid simulation beyond the
  Navier-Stokes equation}

\author{Raoyang Zhang}
\email{raoyang@exa.com}
\author{Xiaowen Shan}
\email{xiaowen@exa.com}
\author{Hudong Chen}
\email{hudong@exa.com}
\affiliation{Exa Corporation, 3 Burlington Woods Drive, Burlington, MA
  01803}

\date{\today}

\begin{abstract}
We present a further theoretical extension to the kinetic theory based
formulation of the lattice Boltzmann method of Shan et al (2006).  In
addition to the higher order projection of the equilibrium
distribution function and a sufficiently accurate Gauss-Hermite
quadrature in the original formulation, a new regularization procedure
is introduced in this paper.  This procedure ensures a consistent
order of accuracy control over the non-equilibrium contributions in
the Galerkin sense.  Using this formulation, we construct a specific
lattice Boltzmann model that accurately incorporates up to the third
order hydrodynamic moments.  Numerical evidences demonstrate that the
extended model overcomes some major defects existed in the
conventionally known lattice Boltzmann models, so that fluid flows at
finite Knudsen number ($\Kn$) can be more quantitatively simulated.
Results from force-driven Poiseuille flow simulations predict the
Knudsen''s minimum and the asymptotic behavior of flow flux at large
$\Kn$.
\end{abstract}


\maketitle

\section{Introduction}

Understanding and simulating fluid flows possessing substantial
non-equilibrium effects pose a long standing challenge to fundamental
statistical physics as well as to many other science and engineering
disciplines~\cite{Reese03,Ho98}.  Due to either rarefaction effects or
small geometric scales, such flows are characterized by a finite
Knudsen number, defined as the ratio between the particle mean free
path, $l$, and the characteristic length, $L$, $\mbox{\sl Kn} =
l/L$. At sufficiently large Knudsen numbers, many of the continuum
assumptions breakdown~\cite{Gad99}.  In particular, the Navier-Stokes
equation and the no-slip boundary condition become inadequate.

Since the Boltzmann equation is valid for describing fluid flows at
any $\Kn$~\cite{Cercignani74}, the conventional approach for
constructing extended hydrodynamic equations for higher $\Kn$ regimes
has been through employing higher order Chapman-Enskog approximations
resulting in, {\em e.g.}, the Burnett and super Burnett equations.
However, this approach encounters both theoretical and practical
difficulties~\cite{Standish99,Agarwal01}.  Alternatively, attempts
have been made to extend the Grad's 13 moment system~\cite{Grad49} by
including contributions of higher kinetic
moments~\cite{Struchtrup02}. One major difficulty has been the
determination of the boundary condition for these moments because only
the lowest few have clear physical meanings.  In addition, due to the
complexity in the resulting equations, application of this approach is
so far limited to simple one-dimensional situations.  Nevertheless,
the moment based formulation offers an valuable insight into the basic
fluid physics for high $\Kn$ flows.

Over the past two decades, the lattice Boltzmann method (LBM) has
developed into an efficient computational fluid dynamic (CFD)
tool~\cite{Chen98}.  Due to its kinetic nature, LBM intrinsically
possesses some essential microscopic physics ingredients and is well
suited for handling more general boundary conditions.  Certain
characteristic phenomena in micro-channel flows were predicted in LBM
simulations at least
qualitatively~\cite{Nie02,Lim02,Li03a,Niu04a,Toschi05,Zhou06,Ansumali05,Ansumali06,Guo06}.
In addition, by introducing a ``stochastic virtual wall collision''
process mimicking effects of free particle streaming in a long
straight channel~\cite{Toschi05}, analytically known asymptotic
behavior at very large $\Kn$ were also produced.  Nevertheless, being
historically developed only to recover fluid physics at the
Navier-Stokes level, the existing LBM schemes used in these studies
possess some well known inaccuracies and numerical artifacts.
Therefore, strictly speaking the schemes are not applicable to high
$\Kn$ flows other than for some rather limited situations. It is
important to develop an LBM method capable of performing accurate and
quantitative simulations of high $\Kn$ flows in general.

Recently, based on the moment expansion formulation~\cite{Shan98}, a
systematic theoretical procedure for extending LBM beyond the
Navier-Stokes hydrodynamics was developed~\cite{Shan06}.  In this
work, we present a specific extended LBM model from this procedure
containing the next order kinetic moments beyond the Navier-Stokes.  A
three-dimensional (3D) realization of this LBM model employs a
39-point Gauss-Hermite quadrature with a sixth order isotropy.  In
addition, a previously reported regularization
procedure~\cite{Chen06,Latt06}, that is fully consistent with the
moment expansion formulation, is incorporated and extended to the
corresponding order.  Simulations performed with the extended LBM have
shown to capture certain characteristic features pertaining to finite
$\Kn$ flows.  There is no empirical models used in the new LBM.

\section{Basic Theoretical Description}

It is theoretically convenient to describe a lattice Boltzmann
equation according to the Hermite expansion
representation~\cite{Shan06}.  The single-particle distribution
functions at a set of particular discrete velocity values, $\{\bxi_a:
a = 1, \cdots, d\}$, are used as the state variables to describe a
fluid system.  The velocity-space discretization is shown to be
equivalent to projecting the distribution function onto a sub-space
spanned by the leading $N$ Hermite ortho-normal basis, denoted by
$\Hs{N}$ hereafter, provided that $\{ \bxi_a \}$ are the abscissas of
a sufficiently accurate Gauss-Hermite quadrature~\cite{Shan98,Shan06}.
Adopting the BGK collision model~\cite{Bhatnagar54}, the discrete
distribution values, $f_a$, satisfy the following equation:
\begin{subequations}
  \begin{eqnarray}
	\label{eq:bgk}
	&& \pp{f_a}t+\bxi_a\cdot\nabla f_a=\Omega_a\\
	&& \Omega_a = -\frac 1\tau \left[f_a-f^{(0)}_a\right] + F_a,
	\quad a = 1, \cdots, d,
  \end{eqnarray}
\end{subequations}
where $\tau$ is a relaxation time, $f^{(0)}_a$ is the truncated Hermite
expansion of the Maxwell-Boltzmann distribution evaluated at $\bxi_a$,
and $F_a$ is the contribution of the body force term.  The truncation
level determines the closeness of the above equation to approximate
the original continuum BGK equation.  A Chapman-Enskog analysis
reveals that the Navier-Stokes equation is recovered when the second
order moment terms are retained. As the higher order terms are
included, physical effects beyond the Navier-Stokes can be captured
systematically~\cite{Shan06}.

In this work we use a specific model of Eq.~(\ref{eq:bgk}) that
consists of moments up to the third order, one order higher than the
Navier-Stokes hydrodynamics in the conventional LBM
models~\cite{Chen98}.  For our present investigation of flows at high
$\Kn$ but low Mach numbers ($\Ma$), here we set the temperature, $T$,
to a constant for simplicity.  Denoting the local fluid density and
velocity by $\rho$ and $\bu$, and defining $u_a = \bxi_a\cdot\bu$ for
brevity, in the dimensionless units in which all velocities are
normalized by the sound speed ({\it i.e.}, $\sqrt{T} = 1$),
$f^{(0)}_a$ takes the following compact form:
\begin{equation}
  \label{eq:maxex}
  f^{(0)}_a = w_a\rho\left[1+u_a + \frac{u_a^2-u^2}2 +
	\frac{u_a\left(u_a^2-3u^2\right)}6\right],
\end{equation}
where $u^2 = \bu\cdot\bu$, and $w_a$ is the quadrature weight
corresponding to the abscissa $\bxi_a$.  The last term inside the
brackets represents the contribution from the third-order kinetic
moments~\cite{Chen97} which was shown to be related to the
velocity-dependent viscosity~\cite{Qian98} but generally neglected in
the conventional lattice Boltzmann models.

According to the previous analysis~\cite{Shan06}, the Gauss-Hermite
quadrature employed for solving a third-order truncated system must be
accurate with the polynomials up to the sixth order.  For ease in
implementing LBM model on Cartesian coordinates, we use the 3D
Cartesian quadrature $E^{39}_{3,7}$ of Ref.~\cite{Shan06}.  Its 2D
projection gives a quadrature $E^{21}_{2,7}$. The abscissas and
weights of both quadratures are provided in Table~\ref{tb:q}.  Both
LBM models can be verified to admit isotropy for tensors of the form
$\sum w_a\bxi_a\cdots\bxi_a$ up to the sixth order instead of fourth
in the conventional LBM models.  Explicitly speaking, recovery of
correct hydrodynamic physics up to the third order requires up to
six-order isotropy conditions as follows~\cite{Chen97}:
\begin{subequations}
  \begin{eqnarray}
	& & \suma w_a = 1\\
    & & \suma w_a\xi_{a,i}\xi_{a,j} = \delta_{ij} \\
    & & \suma w_a\xi_{a,i}\xi_{a,j}\xi_{a,k}\xi_{a,l} 
	= \delta^{(4)}_{ijkl} \\
    & & \suma w_a\xi_{a,i}\xi_{a,j}\xi_{a,k}\xi_{a,l}\xi_{a,m}\xi_{a,n} 
	= \delta^{(6)}_{ijklmn} 
  \end{eqnarray}
\end{subequations}
where the roman subscripts $i$, $j$, $\ldots$, $n$ denote Cartesian
components.  In the above, $\delta_{ij}$ is the Kronecker delta
function, while $\delta^{(4)}_{ijkl}$ and $\delta^{(6)}_{ijklmn}$
represent, respectively, the forth order and the sixth order
generalizations:
\begin{eqnarray}
  \delta^{(4)}_{ijkl} &=& \delta_{ij}\delta_{kl} + 
  \delta_{ik}\delta_{jl} + \delta_{il}\delta_{jk} \nonumber \\
  \delta^{(6)}_{ijklmn} &=& \delta_{ij}\delta^{(4)}_{klmn} + 
  \delta_{ik}\delta^{(4)}_{jlmn} + \delta_{il}\delta^{(4)}_{jkmn} \nonumber \\
  &+& \delta_{im}\delta^{(4)}_{jkln} + \delta_{in}\delta^{(4)}_{jklm}
\end{eqnarray}
Indeed, direct verification shows that these are satisfied by the 3D
39-speed model ($E^{39}_{3,7}$) and its 2D projection
($E^{21}_{2,7}$).  The conventional LBM schemes only satisfy the above
moment isotropy conditions up to the forth order ({\it i.e.}, the
Navier-Stokes order).  It is also important to mention that there
exist a few other lattice Boltzmann models satisfying sixth order
isotropy~\cite{Chen94,Watari04}.  However they contain more discrete
speeds and more complicated coefficients, and are difficult to extend
to even higher orders.

\begin{table}
  \caption{Degree-7 Gauss-Hermite quadratures on Cartesian grid.
  Listed are the number of points in the symmetry group, $p$,
  abscissas, $\bxi_a$, and the weights $w_a$.  Quadratures are named
  by the convention $E^d_{D,n}$ where the superscript $d$ and
  subscripts $D$ and $n$ are respectively the number of abscissas,
  dimension, and degree of algebraic precision.  The subscript $FS$
  denotes permutations with full symmetry.  Note that since all
  velocities are normalized with sound speed, the Cartesian grid
  spacing has a unit velocity of $r=\sqrt{3/2}$.}
  \label{tb:q}
  \begin{ruledtabular}
	\begin{tabular}{c|rrrl}
	  Quadrature     & p & $\bxi_a$              & $w_a$    \\ \hline
	  $E_{3,7}^{39}$ & 1 & $(0, 0, 0)$           & $1/12$   \\
	                 & 6 & $(r, 0, 0)_{FS}$      & $1/12$   \\
	                 & 8 & $(\pm r,\pm r,\pm r)$ & $1/27$   \\
	                 & 6 & $(2r,0, 0)_{FS}$      & $2/135$  \\
	                 &12 & $(2r,2r,0)_{FS}$      & $1/432$  \\
	                 & 6 & $(3r,0, 0)_{FS}$      & $1/1620$ \\ \hline
	  $E_{2,7}^{21}$ & 1 & $(0, 0)$              & $91/324$ \\
	                 & 4 & $(r, 0)_{FS}$         & $1/12$   \\
	                 & 4 & $(\pm r,\pm r)$       & $2/27$   \\
	                 & 4 & $(2r,0)_{FS}$         & $7/360$  \\
	                 & 4 & $(\pm 2r,\pm 2r)$     & $1/432$  \\
	                 & 4 & $(3r,0)_{FS}$         & $1/1620$ \\
	\end{tabular}
  \end{ruledtabular}  
\end{table}

With the Cartesian quadrature above, Eq.~(\ref{eq:bgk}) can be directly
discretized in physical space and time, yielding a standard lattice
Boltzmann equation of the form:
\begin{equation}
  \label{eq:lbgk}
  f_a(\bx+\bxi_a, t+1) = f_a(\bx,t) -
  \frac 1\tau\left[f_a(\bx,t)-f_a^{(0)}\right]+F_a
\end{equation}
As usual, the ``lattice convention'' with unity time increment is used
here.  

\section{The Regularization Procedure}

An LBM computation is generally carried out in two steps: the
streaming step in which $f_a$ at $\bx$ is moved to $\bx+\bxi_a$, and
the collision step in which $f_a(\bx)$ is replaced with the
right-hand-side of Eq.~(\ref{eq:lbgk}).  When viewed as a projection
of the continuum BGK equation into $\Hs{N}$, this dynamic process
introduces an error due to the fact that $f_a$ does not automatically
lie entirely within $\Hs{N}$.  Borrowing the language from spectral
analysis, this is analogous to the aliasing effect.  When the system
is not far from equilibrium, such an error is small and ignorable.  On
the other hand, this error can be resolved via an extension of the
``regularization procedure'' previously designed for improvement in
stability and isotropy~\cite{Chen06,Latt06}.  In terms of the Hermite
expansion interpretation, the regularization procedure is more
concisely described as the following.  We split the post-streaming
distribution into two parts:
\begin{equation}
  f_a = f'_a + f_a^{(0)}
\end{equation}
where $f'_a$ is the deviation from the truncated Maxwellian, or the
{\em non-equilibrium} part of the distribution.  As $f_a^{(0)}$
already lies entirely in the subspace $\Hs{N}$, the projection is to
ensure that the non-equilibrium contribution also lies in the same
subspace for all times, and only needs to be applied to $f'_a$.
Effectively, the projection serves as a filtering (or ``de-aliasing'')
process to ensure the system stay inside the defined subspace $\Hs{N}$
in a Galerkin interpretation.
  
The projection is to convert $f'_a$ to a new distribution
$\widehat{f'_a}$ which lies within the subspace spanned by the first
three Hermite polynomials.  Using the orthogonality relation of the
Hermite polynomials and the Gauss-Hermite quadrature, $\widehat{f'_a}$
is given by the pair of relations:
\begin{subequations}
  \begin{eqnarray}
	\widehat{f'_a} &=& w_a\sum_{n=0}^3\frac 1{n!}\ba{n}\bH{n}(\bxi_a),
	\quad a = 1, \cdots, d,\\
	\ba{n} &=& \suma f'_a\bH{n}(\bxi_a),\quad n = 0, \cdots, 3,
  \end{eqnarray}  
\end{subequations}
where $\bH{n}$ is the standard $n$-th Hermite
polynomial~\cite{Grad49b,Shan98}:
\begin{subequations}
  \begin{eqnarray}
	\bH{0}(\bxi) &=& 1\\
	\bH{1}_i(\bxi) &=& \xi_i\\
	\bH{2}_{ij}(\bxi) &=& \xi_i\xi_j - \delta_{ij}\\
	\bH{3}_{ijk}(\bxi) &=& \xi_i\xi_j\xi_k - \xi_i\delta_{jk} -
	\xi_j\delta_{ik} - \xi_k\delta_{ij},
  \end{eqnarray}
\end{subequations}
and $\ba{n}$ the corresponding Hermite expansion coefficient, both
rank-$n$ tensors.  The first two Hermite coefficients vanish due to
the vanishing contribution from the non-equilibrium distribution to
mass and momentum.  The second and third Hermite coefficients are:
\begin{equation}
  \label{eq:coeff}
  \ba{2} = \suma f'_a\bxi_a\bxi_a,\quad
  \ba{3} = \suma f'_a\bxi_a\bxi_a\bxi_a,
\end{equation}
where $\ba{2}$ is traceless due to the conservation of energy.
Clearly from the above construction or from direct verification,  
the projected distribution $\widehat{f'_a}$ gives the same
second order (momentum) and third order fluxes as the original $f'_a$,
\begin{subequations}
  \begin{eqnarray}
	\suma \widehat{f'_a}\bxi_a\bxi_a &=& \suma f'_a\bxi_a\bxi_a\\
	\suma \widehat{f'_a}\bxi_a\bxi_a\bxi_a &=& \suma f'_a\bxi_a\bxi_a\bxi_a,
  \end{eqnarray}
\end{subequations}
This is an essential step to preserve the required non-equilibrium
properties affecting macroscopic physics.  Furthermore, unlike $f'_a$
in which all higher order moments are in principle present, the
projected distribution $\widehat{f'_a}$ can be shown via the
orthogonality argument to give zero contributions to fluxes higher
than the defined third order above.  Consequently, its physical
implication is rather apparent: The regularization procedure filters
out the higher order non-equilibrium moments that contain strong
discrete artifacts due to the insufficient support of the lattice
basis.

Overall, given the discrete non-equilibrium distribution, its
projection in $\Hs{3}$ is fully specified by:
\begin{equation}
  \label{eq:filterf}
  \widehat{f'_a} = w_a\left[\frac{\bH{2}(\bxi_a)}2\sum_{b=1}^d f'_b\bxi_b\bxi_b
  + \frac{\bH{3}(\bxi_a)}6\sum_{b=1}^d f'_b\bxi_b\bxi_b\bxi_b\right].
\end{equation}
Incorporating the regularization procedure, Eq.~(\ref{eq:lbgk}) is
modified to become
\begin{equation}
  \label{eq:lbgk1}
  f_a(\bx+\bxi_a, t+1) = f_a^{(0)} + \left(1-\frac 1\tau\right)
  \widehat{f'_a} + F_a.
\end{equation}
It is revealing to realize that the right-hand-side represents a
damping operator acting on the ``non-equilibrium'' part of the
dsitribution function.  It is an immediate extension to assign a
different relaxation time to each individual Hermite mode.  Namely we
can recast the collision operator into an equivalent yet slightly more
general form:
\begin{equation}
  \Omega_a = \sum_{b=1}^d{\cal M}_{ab} f'_b,
\end{equation}
where the linear matrix operator takes the following generic form 
\begin{equation}
 \label{eq:matrix}
 {\cal M}_{ab} \equiv \delta_{ab} + \left(1-\frac 1\tau\right)
 w_a\left[\theta_1\frac{\bH{2}(\bxi_a)}2\bxi_b\bxi_b
  + \theta_2\frac{\bH{3}(\bxi_a)}6\bxi_b\bxi_b\bxi_b\right].
\end{equation}
The additional coefficients, $\theta_1$ and $\theta_2$ can have values
between zero and one separately.  Indeed, when $\theta_1 = \theta_2 =
1$, then we recover the result in (\ref{eq:filterf}), the third-order
accurate projection operator.  On the other hand, comparing
Eq.~(\ref{eq:matrix}) with the similar expression in
Ref.~\cite{Chen06}, we find that the latter is in fact a second-order
projection operator ({\it i.e.}, the first term in
Eq.~(\ref{eq:matrix})), or simply $\theta_1 = 1$ and $\theta_2 = 0$.
It is to be emphasized that the correct application of the third-order
projection operator is based on the necessary condition of the third
order Hermite quadrature. Hence it cannot be realized via conventional
low order LBM such as the popular D3Q19 (or D2Q9).

The explicit form of the body force term comes directly from the
Hermite expansion of the continuum BGK equation~\cite{Martys98,Shan06}.  
Up to the third order, it can be expressed as:
\begin{eqnarray}
  \label{eq:force}
  F_a &=& w_a\rho\bg\cdot\left(\bxi_a+u_a\bxi_a-\bu\right)\nonumber\\
  &+&\frac 12w_a\left[\ba{2}+\rho\bu\bu\right]:
  \left[g_a\bH{2}(\bxi_a)-2\bg\bxi_a\right].
\end{eqnarray}
To be noted here is that, whereas the first term is entirely due to
the equilibrium part of the distribution, the term related to $\ba{2}$
are contributions from the non-equilibrium part.  To our knowledge,
the non-equilibrium contribution in the body-force has not been
explicitly considered in the existing LBM literature.  Nevertheless,
although it is
expected to play an important role at large $\Kn$, at moderate $\Kn$
($\leq 1$), no significant effects due to non-equilibrium contribution
are observed in the numerical experiments in the present work.

\begin{figure}
  \includegraphics[viewport=0 50 730 550,width=\columnwidth]{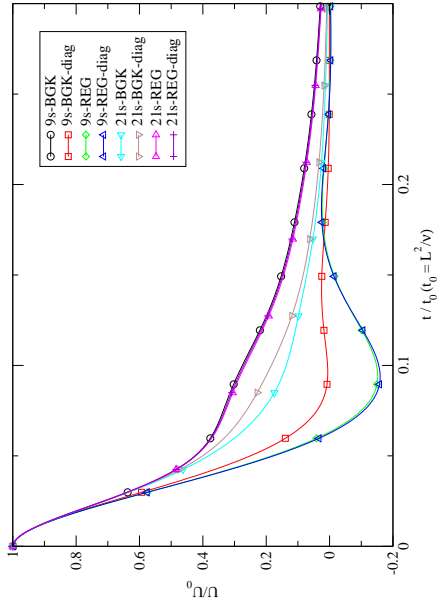}
  \caption{(Color online) Peak velocity of a decaying shear waves as
  simulated by the 9-state (9s) and the 21-state (21s) standard BGK
  and regularized (REG) models.  For each model, simulation is carried
  out with the wave vector aligned with either the lattice links or
  their diagonals.  The latter is denoted by the post-fix ``diag''}
  \label{fg:shere}
\end{figure}

\section{Numerical Results}

\subsection{Shear wave decay}

The first series of numerical simulations performed are for the
benchmark problem of 2D sinusoidal shear wave decay.  This set of
tests is to evaluate impact from the increased order of accuracy and
from the regularization procedure on the resulting isotropy.
Obviously, a numerical artifact free wave decay should not depend on
its relative orientation with respect to the underlying lattice
orientation.  The lattice orientation dependent artifact has often
plagued discrete fluid models especially at finite Knudsen number when
non-equilibrium effects are significant.  For this purpose, we have
defined two sets of simulations.  In the first set, a sinusoidal wave
with a wavelength ($L$) of 128 grid spacing is simulated on a
$128\times 128$ periodic domain.  The initial velocity field is given
by $u_x=u_0\sin(y/2\pi L)$ and $u_y = 0$.  The wave vector is aligned
with the lattice.  In the second set, the same sinusoidal wave is
rotated by 45 degrees from the original orientation and simulated on a
matching periodic domain size of $181(=128\sqrt{2})\times 181$.  The
Knudsen number, defined as $\Kn = 2\tau c_s/L$, is chosen to be $0.2$,
where $\tau$ is the relaxation time and $c_s$ the sound speed. These
two sets of simulations were conducted using four representative
models: the popular 2D 9-state (9-s) model (D3Q9), and the present 2D
21-state (21-s) model based on $E^{21}_{2,7}$, both with and without
the regularization process.  Note, the 2D 9-state model only admits a
second-order regularization projection as discussed in the preceding
section. In discussions hereafter we shall refer the models without
the regularization as the BGK models, and the ones with regularization
the REG models.

In Fig.~\ref{fg:shere}, the dimensionless peak velocity magnitude,
normalized by its initial value and measured at the $1/4$ width of
wavelength, is plotted against the non-dimensionalized time normalized
by the characteristic decay time $t_0 = \lambda^2/ \nu$, where $\nu =
c_s^2\left(\tau - 1/2\right)$ is kinematic viscosity.  As one can
easily notice, the decay rate of the shear wave for the 9-s BGK model
is substantially different between the lattice oriented setup and the
45 degree one, exhibiting a strong dependence on the orientation of
the wave vector with respect to the lattice. This indicates a strong
anisotropy of the model at this $\Kn$.  This is consistent to our
expectation that the 9-s BGK model was originally formulated to only
recover the Navier-Stokes hydrodynamics ({\it i.e.}, at vanishing
$\Kn$).  Interestingly this anisotropy is essentially eliminated by
the second-order regularization procedure in the resulting 9-s REG
model.  On the other hand, the amplitude of the shear wave exhibits a
strong oscillatory behavior in addition to the exponential decay,
implying a greater than physical ballistic effect.  These may be
explained as the following: The non-equilibrium part of the
post-streaming distribution contains contributions from {\em in
principle} all moments, which are highly anisotropic due to inadequate
(only up to the second-order moment) symmetry in the underlying
discrete model.  The regularization procedure filters out all the
higher than the second order moment contributions, yielding an
isotropic behavior supported by the given lattice.  On the other hand,
the higher moments are critical at large $\Kn$.  Therefore, though
isotropic, the 9-s REG model still should not be expected to show
satisfactory physical results at high Knudsen numbers.  For the 21-s
BGK model, an anisotropic behavior is also very visible, though at a
much smaller extent.  This may be attributed to the residual
anisotropy in the moments higher than the third order.  Again, the
anisotropy behavior is completely removed once the regularization
procedure is applied in the 21-s REG model, as shown from the totally
overlapped curves between the lattice oriented and the diagonally
oriented simulations.  It is also noticeable that the decay history
shows a much reduced oscillatory behavior in the 21-s REG model.
Because of its correct realization of the third order moment flux, we
expect the result is more accurate at this Knudsen number value.  It
is also curious to observe that the decay of the ``lattice aligned''
result from 9s-BGK is surprisingly close to that of 21-s REG model at
this $\Kn$.  This is likely to be a mere coincidence.

\subsection{Finite Knudsen number channel flows}

Using the same four models, we subsequently carried out simulations of
the force-driven Poiseuille flow for a range of Knudsen numbers.  In
order to identify the impact in accuracy in the resulting lattice
Boltzmann models as oppose to the effects from various boundary
conditions, here a standard half-way bounce-back boundary condition is
used in the cross-channel ($y$) direction.  Furthermore, since we are
not interested, for the present study, in any physical phenomenon
associated with stream-wise variations, a periodic boundary condition
is used with only four grid points in the stream-wise ($x$) direction.
In the cross stream ($y$) direction, two different resolutions, $L=40$
and $80$ are both tried to ensure sufficient resolution independence.
The Knudsen number is defined as $\Kn = \nu/(c_s L)$.  The flow is
driven by a constant gravity force, $\bg$, pointing in the positive
$x$ direction.  The magnitude of the force is set to $8\nu U_0/L^2$,
which would give rise to a parabolic velocity profile with a peak
velocity of $U_0$ in the vanishing $\Kn$ limit.  For consistence, a
modified definition of fluid velocity, $\bu \rightarrow \bu +\bg /2$,
is used in $f_a^{(0)}$.  Since the LBM models presently used here are
all assuming constant temperature, to enforce an incompressible
behavior with negligible thermodynamic effect throughout the simulated
$\Kn$ range, we choose a sufficiently small value of $U_0$,
corresponding to the nominal Mach numbers of $\Ma (= U_0/c_s) \sim
1.46 \times 10^{-6}$ and $2.92 \times 10^{-7}$, and verified that our
results are independent of $\Ma$.  The actual resulting fluid velocity
in these simulations can achieve values much higher than $U_0$ at
higher $\Kn$.

\begin{figure}
  \includegraphics[viewport=0 50 730 550,width=\columnwidth]{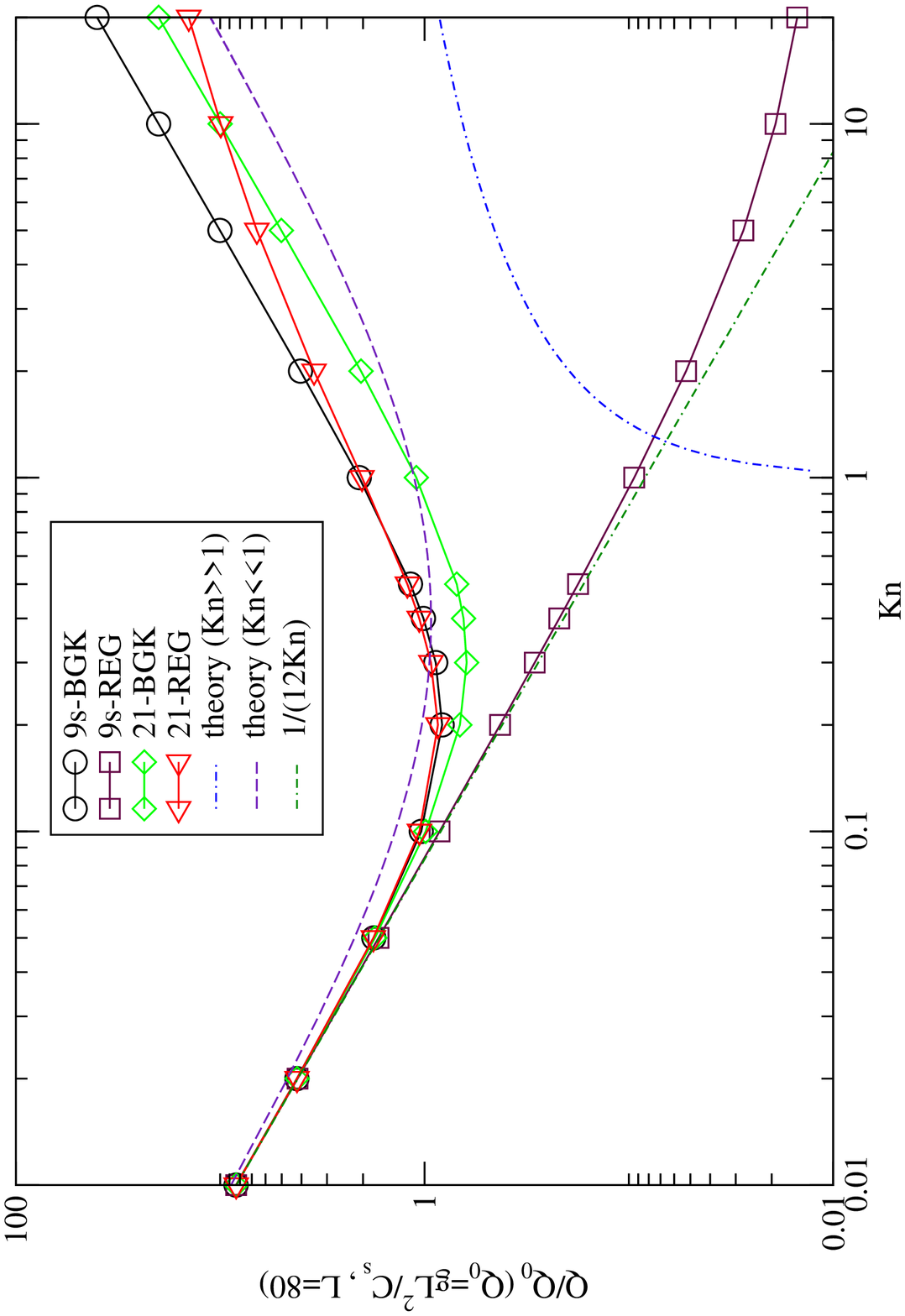}
  \caption{(Color online) This figure shows the Knudsen paradox with
	resolution 40, $\Ma = 1.46 \times 10^{-6}$.  The $\Kn$ which has
	the minimum $Q$ is about 0.2 for 21s-REG, 0.3 for 21s-BGK and 0.2
	for 9s-BGK. The theoretical results are that of
	Cercignani~\cite{Cercignani74}.}
  \label{fg:flux}
\end{figure}

\begin{figure}
  \includegraphics[viewport=0 50 730 550,width=\columnwidth]{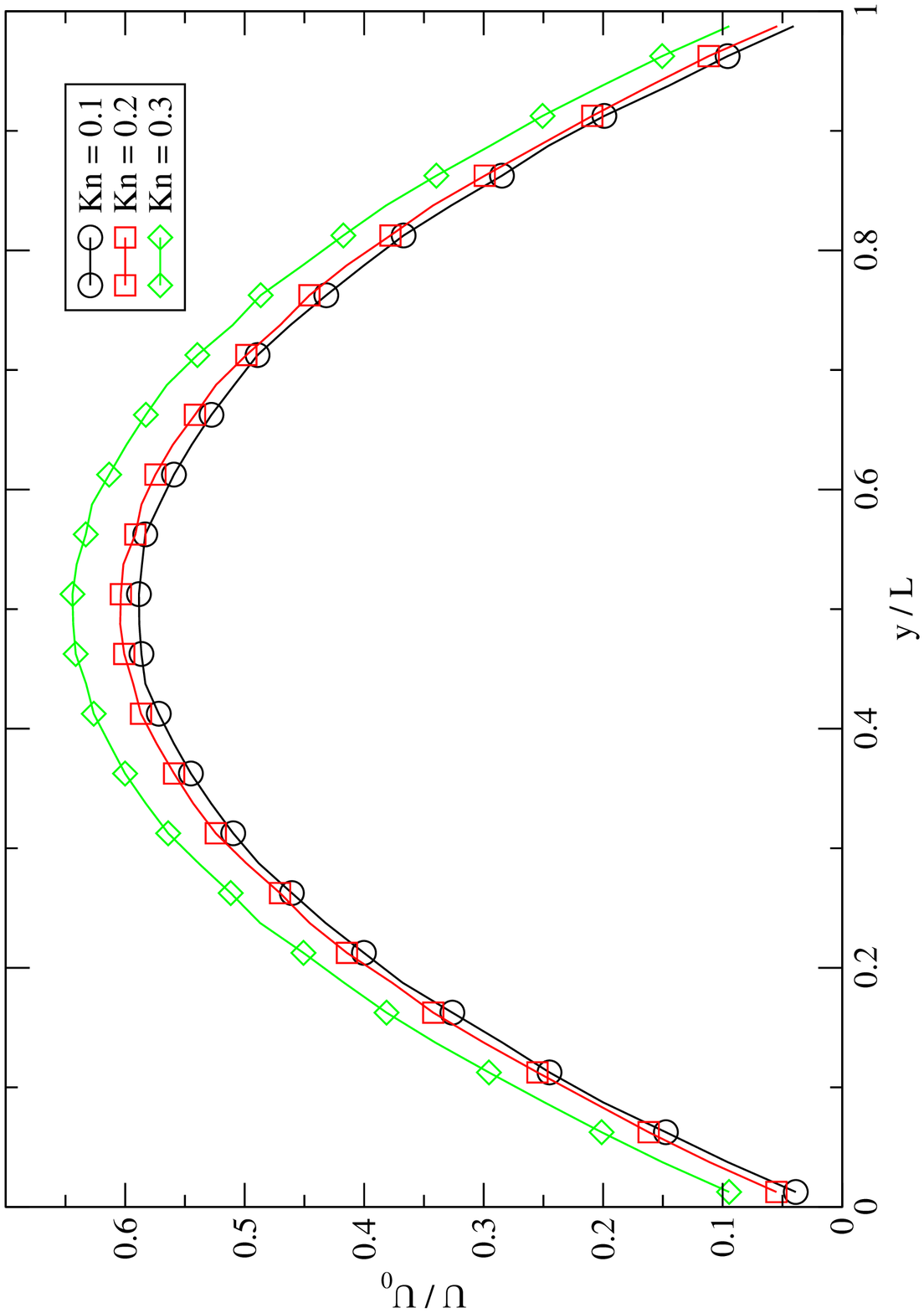}
  \caption{(Color online) This figure shows the normalized velocity
	profiles at resolution 40 and $\Kn = 0.1$, 0.2, and 0.5.  Notice
	the 9-s REG model exhibits no visible slip velocity for both small
	and large $\Kn$.}
  \label{fg:profile1}
\end{figure}

\begin{figure}
  \includegraphics[viewport=0 50 730 550,width=\columnwidth]{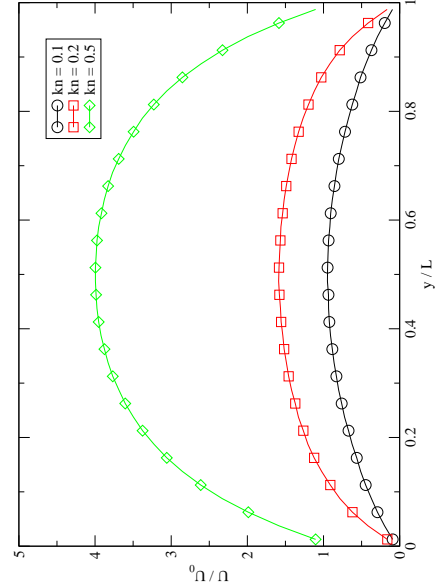}
  \caption{(Color online) This figure shows the normalized velocity
	profiles at resolution 40 and $\Kn = 0.1$, 0.2, and 0.5.  Notice
	the 21-s REG model exhibits substantial slip velocity at large
	$\Kn$.}
  \label{fg:profile2}
\end{figure}

Plotted in Fig.~\ref{fg:flux} is the non-dimensionalized mass flux, $Q
\equiv \sum_{y = 0}^L u_x(y)/Q_0$, as a function of $\Kn$ in the final
steady state of the simulations.  Here $Q_0 = gL^2/c_s$.  For
comparison we also include two analytical asymptotic
solutions~\cite{Cercignani74} for both small and very large $\Kn$.  To
be noted first is that at the vanishing $\Kn$ limit, all simulation
results agree with each other as well as with the analytical solution.
This confirms that all these LBM schemes recover the correct
hydrodynamic behavior at vanishing $\Kn$, {\it i.e.}, the
Navier-Stokes limit.  Also plotted is the exact Navier-Stokes solution
of $Q = 1/(12\Kn )$, a well understood monotonically decreasing line
with no minimum.  At higher $\Kn$, the 9-s BGK model exhibits a
Knudsen's minimum while over-estimates the flux according to some
previously published reports~\cite{Toschi05}. However, by filtering
out moment contributions higher than the second order, such a
phenomenon is completely disappeared from the result of 9-s REG,
yielding a purely monotonically decreasing behavior.  This is a rather
interesting but not entirely surprising result.  Once again, the
regularization process enforces the system to be confined within the
second-order Hermite moment space, while all higher order
non-equilibrium contributions including both the numerical artifacts
and the physical ones, responsible for the finite Knudsen phenomena,
are filtered out.  Consequently, only the Navier-Stokes order effects
are preserved in the 9-s REG model no matter the degree of
non-equilibrium at finite $\Kn$.  To be further noticed is the impact
of the second-order regularization on the near wall properties. In
particular, Fig.~\ref{fg:profile1} shows that 9-s REG gives vanishing
slip velocity at the wall for a range of $\Kn$ values ($\Kn = 0.1$,
0.2, and 0.5).  This suggests that a bounce-back boundary process is
sufficient to realize a no-slip condition once a Navier-Stokes order
dynamics in the model equation is enforced.  This is yet another
confirmation as to why the resulting $Q$ from 9-s REG model lies very
close to the exact Navier-Stokes theoretical curve up to significantly
high $\Kn$ values.  In comparison, all the other three LBM models show
finite slip velocity values, due to the previous discussed reason that
they all contain higher than the second-order non-equilibrium
contributions.  Once again, one must remember that the higher order
effects in 9-s BGK (and to a lesser extent the 21-s BGK) have
substantial lattice discrete artifacts. To be emphasized is the 21-s
REG model: Due to the regularization procedure, only the third-order
non-equilibrium moment contribution is preserved.  On the other hand,
because it has numerically shown to capture the Knudsen minimum
phenomenon, correct incorporation of the third-order moment physics is
thus essential for accurately simulating some key flow physics beyond
the Navier-Stokes.  We also wish to emphasize that all these
differences are due to the intrinsic natures in these LBM models, and
has nothing to do with spatial and temporal resolutions. As a
comparison, we plot in Fig.~\ref{fg:profile2} velocity profiles
generated from the 21-s REG model.

There are a number of ways that gravity force can be included in LB
equations.  One can either treat the gravity force outside the
collision operator as an external body force term as given by
Eq.~(\ref{eq:force}) or by applying a local momentum/velocity shift
inside the equilibrium distribution~\cite{Shan93}.  With regular BGK
model, for all $\Kn$, no difference in results are observed when the
gravity force is applied in different ways.  With the regularization
procedure and sufficient isotropy however, some differences are
observed at finite ($> 1$) $\Kn$.  Generally speaking, applying the
body force via Eq.~(\ref{eq:force}) tends to predict higher flux then
via momentum/velocity shifting.  For the results shown in
Fig.~\ref{fg:flux}, the velocity shift applied in $f_a^{(0)}$ is $1/2
{\bf g}$.  The other half of the gravity force is applied as an
external body force term~(\ref{eq:force}) (Cf.,~\cite{Martys98}).

The results from both the 21-s BGK and the 21-s REG models predict a
Knudsen minimum which resembles that of the 9-s BGK except with
reduced over-estimations at higher $\Kn$.  What is interesting, and
requires further understanding, is that the flux behavior predicted by
the 21-s REG exhibits a reversal of curvature at higher $\Kn$,
resembling the analytical asymptotic solution of
Cercignani~\cite{Cercignani74}.

The qualitative differences seen from these four models suggest that
contributions from moments beyond second order are essential for
capturing fundamental physical effects at high-$\Kn$.  Although the
high-order moments are implicitly contained in the second-order BGK
model, its dynamics is highly contaminated with numerical lattice
artifacts.  In contrast, by incorporating the high-order moments
explicitly and systematically with the regularization, flows at these
$\Kn$ values can indeed be captured by the extended LBM model.

\section{Conclusion}

In summary, the kinetic based representation offers a well-posed
approach in formulation of computational models for performing
efficient and quantitative numerical simulations of fluid flows at
finite $\Kn$.  In this work we present a specific extended LBM model
that accurately incorporates the physical contributions of kinetic
moments up to the third order in Hermite expansion space.  The new
regularization procedure presented in this paper ensures that both the
equilibrium and the non-equilibrium effects are confined in the
accurately supported truncated subspace at all times so that the
unphysical artifacts are filtered out in the dynamic process. This
resulting LBM model is robust and highly efficient.  Because of its
accurate inclusion of the essential third order contributions, this
model is demonstrated to be capable of quantitatively capturing
certain fundamental flow physics properties at finite Knudsen numbers.
This is accomplished without imposing any empirical models.
Furthermore, because of the removal of discrete anisotropy, it is also
clear that the new LBM model is not limited to specific
uni-directional channel flows, nor it is only applicable for lattice
aligned orientations.

Nevertheless, a number of issues are awaiting further studies.  For
even higher $\Kn$ ($\sim 10$), one should expect moment contributions
higher than the third order to become physically important.  This is
straightforward to include via the systematic
formulation~\cite{Shan06} together with the regularization procedure
described here.  The issue of boundary condition is also of crucial
importance~\cite{Ginzburg94,Ginzburg96,Chen98a}.  Even though, as
demonstrated in this paper that the realization of the essential
slip-velocity effect and the asymptotic behavior is attributed {\em to
a significant extent} to the third order and higher moment
contributions in the intrinsic LBM dynamic model itself.  As reported
in some previous works~\cite{Niu04a,Toschi05}, the kinetic boundary
condition of Ansumali and Karlin~\cite{Ansumali02a} has led to
substantial improvements at the Navier-Stokes level micro-channel flow
simulations.  Boundary condition itself serves as an effective
collision so that it also modifies the degree of the non-equilibrium
distribution.  Specifically, the well known Maxwell boundary condition
(in which the particle distribution function is assumed to be at
equilibrium) should be expected give different effect at high $\Kn$
compared with that of the bounce-back used in the present study. The
latter boundary condition preserves non-equilibrium contributions at
all orders.  As a consequence, we suspect that the finite slip
phenomenon is likely to still be over predicted in our current
simulations with the bounce-back boundary condition than what actually
would occur in reality.  Many more detailed and further investigations
particularly pertaining channel flows at finite Knudsen number are
certainly extremely important in the future studies.

Thermodynamic effect is also expected to become important at finite
$\Kn$ when Mach number is not negligibly small. Some distinctive
phenomena that is substantial and characteristic only at finite
Knudsen numbers with sufficiently large Mach numbers~\cite{Xu03}.  The
work of Xu ~\cite{Xu03} demonstrates the importance of an accurate
formulation of higher than the Navier-Stokes order thermodynamic
effect at finite $\Kn$. Theoretically, this additional property is
associated with the so called super-Burnett effect in the more
conventional language~\cite{Xu03}, or the forth-order moment
contribution in the Hermite expansion and can be incorporated in a
further extended LBM model~\cite{Shan06}.  The present third-order
model is thermodynamically consistent, but only at the Navier-Stokes
level~\cite{Shan06}.  Another interesting point to mention is that
both the 21-s BGK or the 21-s REG can allow an expanded equilibrium
distribution form including terms up to the fourth power in fluid
velocity (as opposed to the square power in 9-s BGK), for that the
correct equilibrium energy flux tensor is still preserved.  Including
the forth power terms immediately results in a desirable
positive-definite distribution for the zero-velocity state at all Mach
number values.

The work is supported in part by the National Science Foundation.


\begin{thebibliography}{35}
\expandafter\ifx\csname natexlab\endcsname\relax\def\natexlab#1{#1}\fi
\expandafter\ifx\csname bibnamefont\endcsname\relax
  \def\bibnamefont#1{#1}\fi
\expandafter\ifx\csname bibfnamefont\endcsname\relax
  \def\bibfnamefont#1{#1}\fi
\expandafter\ifx\csname citenamefont\endcsname\relax
  \def\citenamefont#1{#1}\fi
\expandafter\ifx\csname url\endcsname\relax
  \def\url#1{\texttt{#1}}\fi
\expandafter\ifx\csname urlprefix\endcsname\relax\def\urlprefix{URL }\fi
\providecommand{\bibinfo}[2]{#2}
\providecommand{\eprint}[2][]{\url{#2}}

\bibitem[{\citenamefont{Reese et~al.}(2003)\citenamefont{Reese, Gallis, and
  Lockerby}}]{Reese03}
\bibinfo{author}{\bibfnamefont{J.~M.} \bibnamefont{Reese}},
  \bibinfo{author}{\bibfnamefont{M.~A.} \bibnamefont{Gallis}},
  \bibnamefont{and} \bibinfo{author}{\bibfnamefont{D.~A.}
  \bibnamefont{Lockerby}}, \bibinfo{journal}{Phil. Trans. Roy. Soc. London A}
  \textbf{\bibinfo{volume}{361}}, \bibinfo{pages}{2967} (\bibinfo{year}{2003}).

\bibitem[{\citenamefont{Ho and Tai}(1998)}]{Ho98}
\bibinfo{author}{\bibfnamefont{C.-M.} \bibnamefont{Ho}} \bibnamefont{and}
  \bibinfo{author}{\bibfnamefont{Y.-C.} \bibnamefont{Tai}},
  \bibinfo{journal}{Annu. Rev. Fluid Mech.} \textbf{\bibinfo{volume}{30}},
  \bibinfo{pages}{579} (\bibinfo{year}{1998}).

\bibitem[{\citenamefont{{Gad-el-Hak}}(1999)}]{Gad99}
\bibinfo{author}{\bibfnamefont{M.}~\bibnamefont{{Gad-el-Hak}}},
  \bibinfo{journal}{J. Fluids Eng.} \textbf{\bibinfo{volume}{121}},
  \bibinfo{pages}{5} (\bibinfo{year}{1999}).

\bibitem[{\citenamefont{Cercignani}(1974)}]{Cercignani74}
\bibinfo{author}{\bibfnamefont{C.}~\bibnamefont{Cercignani}},
  \emph{\bibinfo{title}{Theory and Application of the Boltzmann Equation}}
  (\bibinfo{publisher}{Scottish Academic Press Ltd.}, \bibinfo{address}{New
  York}, \bibinfo{year}{1974}).

\bibitem[{\citenamefont{Standish}(1999)}]{Standish99}
\bibinfo{author}{\bibfnamefont{R.~K.} \bibnamefont{Standish}},
  \bibinfo{journal}{Phys. Rev. E} \textbf{\bibinfo{volume}{60}},
  \bibinfo{pages}{5175} (\bibinfo{year}{1999}).

\bibitem[{\citenamefont{Agarwal et~al.}(2001)\citenamefont{Agarwal, Yun, and
  Balakrishnan}}]{Agarwal01}
\bibinfo{author}{\bibfnamefont{R.~K.} \bibnamefont{Agarwal}},
  \bibinfo{author}{\bibfnamefont{K.-Y.} \bibnamefont{Yun}}, \bibnamefont{and}
  \bibinfo{author}{\bibfnamefont{R.}~\bibnamefont{Balakrishnan}},
  \bibinfo{journal}{Phys. Fluids} \textbf{\bibinfo{volume}{13}},
  \bibinfo{pages}{3061} (\bibinfo{year}{2001}).

\bibitem[{\citenamefont{Grad}(1949{\natexlab{a}})}]{Grad49}
\bibinfo{author}{\bibfnamefont{H.}~\bibnamefont{Grad}},
  \bibinfo{journal}{Commun. Pure Appl. Math.} \textbf{\bibinfo{volume}{2}},
  \bibinfo{pages}{331} (\bibinfo{year}{1949}{\natexlab{a}}).

\bibitem[{\citenamefont{Struchtrup}(2002)}]{Struchtrup02}
\bibinfo{author}{\bibfnamefont{H.}~\bibnamefont{Struchtrup}},
  \bibinfo{journal}{Phys. Rev. E} \textbf{\bibinfo{volume}{65}},
  \bibinfo{pages}{041204} (\bibinfo{year}{2002}).

\bibitem[{\citenamefont{Chen and Doolen}(1998)}]{Chen98}
\bibinfo{author}{\bibfnamefont{S.}~\bibnamefont{Chen}} \bibnamefont{and}
  \bibinfo{author}{\bibfnamefont{G.}~\bibnamefont{Doolen}},
  \bibinfo{journal}{Ann. Rev. Fluid Mech.} \textbf{\bibinfo{volume}{30}},
  \bibinfo{pages}{329} (\bibinfo{year}{1998}).

\bibitem[{\citenamefont{Nie et~al.}(2002)\citenamefont{Nie, Doolen, and
  Chen}}]{Nie02}
\bibinfo{author}{\bibfnamefont{X.}~\bibnamefont{Nie}},
  \bibinfo{author}{\bibfnamefont{G.~D.} \bibnamefont{Doolen}},
  \bibnamefont{and} \bibinfo{author}{\bibfnamefont{S.}~\bibnamefont{Chen}},
  \bibinfo{journal}{J. Stat. Phys.} \textbf{\bibinfo{volume}{107}},
  \bibinfo{pages}{279} (\bibinfo{year}{2002}).

\bibitem[{\citenamefont{Lim et~al.}(2002)\citenamefont{Lim, Shu, Niu, and
  Chew}}]{Lim02}
\bibinfo{author}{\bibfnamefont{C.~Y.} \bibnamefont{Lim}},
  \bibinfo{author}{\bibfnamefont{C.}~\bibnamefont{Shu}},
  \bibinfo{author}{\bibfnamefont{X.~D.} \bibnamefont{Niu}}, \bibnamefont{and}
  \bibinfo{author}{\bibfnamefont{Y.~T.} \bibnamefont{Chew}},
  \bibinfo{journal}{Phys. Fluids} \textbf{\bibinfo{volume}{14}},
  \bibinfo{pages}{2299} (\bibinfo{year}{2002}).

\bibitem[{\citenamefont{Li and Kwok}(2003)}]{Li03a}
\bibinfo{author}{\bibfnamefont{B.}~\bibnamefont{Li}} \bibnamefont{and}
  \bibinfo{author}{\bibfnamefont{D.~Y.} \bibnamefont{Kwok}},
  \bibinfo{journal}{Phys. Rev. Lett.} \textbf{\bibinfo{volume}{90}},
  \bibinfo{pages}{124502} (\bibinfo{year}{2003}).

\bibitem[{\citenamefont{Niu et~al.}(2004)\citenamefont{Niu, Shu, and
  Chew}}]{Niu04a}
\bibinfo{author}{\bibfnamefont{X.~D.} \bibnamefont{Niu}},
  \bibinfo{author}{\bibfnamefont{C.}~\bibnamefont{Shu}}, \bibnamefont{and}
  \bibinfo{author}{\bibfnamefont{Y.~T.} \bibnamefont{Chew}},
  \bibinfo{journal}{Europhys. Lett.} \textbf{\bibinfo{volume}{67}},
  \bibinfo{pages}{600} (\bibinfo{year}{2004}).

\bibitem[{\citenamefont{Toschi and Succi}(2005)}]{Toschi05}
\bibinfo{author}{\bibfnamefont{F.}~\bibnamefont{Toschi}} \bibnamefont{and}
  \bibinfo{author}{\bibfnamefont{S.}~\bibnamefont{Succi}},
  \bibinfo{journal}{Europhys. Lett.} \textbf{\bibinfo{volume}{69}},
  \bibinfo{pages}{549} (\bibinfo{year}{2005}).

\bibitem[{\citenamefont{Zhou et~al.}(2006)\citenamefont{Zhou, Zhang,
  Staroselsky, Chen, Kim, and Jhon}}]{Zhou06}
\bibinfo{author}{\bibfnamefont{Y.}~\bibnamefont{Zhou}},
  \bibinfo{author}{\bibfnamefont{R.}~\bibnamefont{Zhang}},
  \bibinfo{author}{\bibfnamefont{I.}~\bibnamefont{Staroselsky}},
  \bibinfo{author}{\bibfnamefont{H.}~\bibnamefont{Chen}},
  \bibinfo{author}{\bibfnamefont{W.}~\bibnamefont{Kim}}, \bibnamefont{and}
  \bibinfo{author}{\bibfnamefont{M.~S.} \bibnamefont{Jhon}},
  \bibinfo{journal}{Physica {A}} \textbf{\bibinfo{volume}{362}},
  \bibinfo{pages}{68} (\bibinfo{year}{2006}).

\bibitem[{\citenamefont{Ansumali and Karlin}(2005)}]{Ansumali05}
\bibinfo{author}{\bibfnamefont{S.}~\bibnamefont{Ansumali}} \bibnamefont{and}
  \bibinfo{author}{\bibfnamefont{I.~V.} \bibnamefont{Karlin}},
  \bibinfo{journal}{Phys. Rev. Lett.} \textbf{\bibinfo{volume}{95}},
  \bibinfo{pages}{260605} (\bibinfo{year}{2005}).

\bibitem[{\citenamefont{Ansumali et~al.}(2006)\citenamefont{Ansumali, Karlin,
  Frouzakis, and Boulouchos}}]{Ansumali06}
\bibinfo{author}{\bibfnamefont{S.}~\bibnamefont{Ansumali}},
  \bibinfo{author}{\bibfnamefont{I.}~\bibnamefont{Karlin}},
  \bibinfo{author}{\bibfnamefont{C.~E.} \bibnamefont{Frouzakis}},
  \bibnamefont{and}
  \bibinfo{author}{\bibfnamefont{K.}~\bibnamefont{Boulouchos}},
  \bibinfo{journal}{Physica A} \textbf{\bibinfo{volume}{359}},
  \bibinfo{pages}{289} (\bibinfo{year}{2006}).

\bibitem[{\citenamefont{Guo et~al.}(2006)\citenamefont{Guo, Zhao, and
  Shi}}]{Guo06}
\bibinfo{author}{\bibfnamefont{Z.}~\bibnamefont{Guo}},
  \bibinfo{author}{\bibfnamefont{T.~S.} \bibnamefont{Zhao}}, \bibnamefont{and}
  \bibinfo{author}{\bibfnamefont{Y.}~\bibnamefont{Shi}}, \bibinfo{journal}{J.
  Appl. Phys.} \textbf{\bibinfo{volume}{99}}, \bibinfo{pages}{074903}
  (\bibinfo{year}{2006}).

\bibitem[{\citenamefont{Shan and He}(1998)}]{Shan98}
\bibinfo{author}{\bibfnamefont{X.}~\bibnamefont{Shan}} \bibnamefont{and}
  \bibinfo{author}{\bibfnamefont{X.}~\bibnamefont{He}}, \bibinfo{journal}{Phys.
  Rev. Lett.} \textbf{\bibinfo{volume}{80}}, \bibinfo{pages}{65}
  (\bibinfo{year}{1998}).

\bibitem[{\citenamefont{Shan et~al.}(2006)\citenamefont{Shan, Yuan, and
  Chen}}]{Shan06}
\bibinfo{author}{\bibfnamefont{X.}~\bibnamefont{Shan}},
  \bibinfo{author}{\bibfnamefont{X.-F.} \bibnamefont{Yuan}}, \bibnamefont{and}
  \bibinfo{author}{\bibfnamefont{H.}~\bibnamefont{Chen}}, \bibinfo{journal}{J.
  Fluid Mech.} \textbf{\bibinfo{volume}{550}}, \bibinfo{pages}{413}
  (\bibinfo{year}{2006}).

\bibitem[{\citenamefont{Chen et~al.}(2006)\citenamefont{Chen, Zhang,
  Staroselsky, and Jhon}}]{Chen06}
\bibinfo{author}{\bibfnamefont{H.}~\bibnamefont{Chen}},
  \bibinfo{author}{\bibfnamefont{R.}~\bibnamefont{Zhang}},
  \bibinfo{author}{\bibfnamefont{I.}~\bibnamefont{Staroselsky}},
  \bibnamefont{and} \bibinfo{author}{\bibfnamefont{M.}~\bibnamefont{Jhon}},
  \bibinfo{journal}{Physica {A}} \textbf{\bibinfo{volume}{362}},
  \bibinfo{pages}{125} (\bibinfo{year}{2006}).

\bibitem[{\citenamefont{Latt and Chopard}(2006)}]{Latt06}
\bibinfo{author}{\bibfnamefont{J.}~\bibnamefont{Latt}} \bibnamefont{and}
  \bibinfo{author}{\bibfnamefont{B.}~\bibnamefont{Chopard}},
  \bibinfo{journal}{Mathematics and Computers in Simulations}
  \textbf{\bibinfo{volume}{72}} (\bibinfo{year}{2006}).

\bibitem[{\citenamefont{Bhatnagar et~al.}(1954)\citenamefont{Bhatnagar, Gross,
  and Krook}}]{Bhatnagar54}
\bibinfo{author}{\bibfnamefont{P.~L.} \bibnamefont{Bhatnagar}},
  \bibinfo{author}{\bibfnamefont{E.~P.} \bibnamefont{Gross}}, \bibnamefont{and}
  \bibinfo{author}{\bibfnamefont{M.}~\bibnamefont{Krook}},
  \bibinfo{journal}{Phys. Rev.} \textbf{\bibinfo{volume}{94}},
  \bibinfo{pages}{511} (\bibinfo{year}{1954}).

\bibitem[{\citenamefont{Chen et~al.}(1997)\citenamefont{Chen, Teixeira, and
  Molvig}}]{Chen97}
\bibinfo{author}{\bibfnamefont{H.}~\bibnamefont{Chen}},
  \bibinfo{author}{\bibfnamefont{C.}~\bibnamefont{Teixeira}}, \bibnamefont{and}
  \bibinfo{author}{\bibfnamefont{K.}~\bibnamefont{Molvig}},
  \bibinfo{journal}{Intl. J. Mod. Phys. C} \textbf{\bibinfo{volume}{8}},
  \bibinfo{pages}{675} (\bibinfo{year}{1997}).

\bibitem[{\citenamefont{Qian and Zhou}(1998)}]{Qian98}
\bibinfo{author}{\bibfnamefont{Y.-H.} \bibnamefont{Qian}} \bibnamefont{and}
  \bibinfo{author}{\bibfnamefont{Y.}~\bibnamefont{Zhou}},
  \bibinfo{journal}{Europhys. Lett.} \textbf{\bibinfo{volume}{42}},
  \bibinfo{pages}{359} (\bibinfo{year}{1998}).

\bibitem[{\citenamefont{Chen et~al.}(1994)\citenamefont{Chen, Ohashi, and
  Akiyama}}]{Chen94}
\bibinfo{author}{\bibfnamefont{Y.}~\bibnamefont{Chen}},
  \bibinfo{author}{\bibfnamefont{H.}~\bibnamefont{Ohashi}}, \bibnamefont{and}
  \bibinfo{author}{\bibfnamefont{M.}~\bibnamefont{Akiyama}},
  \bibinfo{journal}{Phys. Rev. E} \textbf{\bibinfo{volume}{50}},
  \bibinfo{pages}{2776} (\bibinfo{year}{1994}).

\bibitem[{\citenamefont{Watari and Tsutahara}(2004)}]{Watari04}
\bibinfo{author}{\bibfnamefont{M.}~\bibnamefont{Watari}} \bibnamefont{and}
  \bibinfo{author}{\bibfnamefont{M.}~\bibnamefont{Tsutahara}},
  \bibinfo{journal}{Phys. Rev. E} \textbf{\bibinfo{volume}{70}},
  \bibinfo{pages}{016703} (\bibinfo{year}{2004}).

\bibitem[{\citenamefont{Grad}(1949{\natexlab{b}})}]{Grad49b}
\bibinfo{author}{\bibfnamefont{H.}~\bibnamefont{Grad}},
  \bibinfo{journal}{Commun. Pure Appl. Math.} \textbf{\bibinfo{volume}{2}},
  \bibinfo{pages}{325} (\bibinfo{year}{1949}{\natexlab{b}}).

\bibitem[{\citenamefont{Martys et~al.}(1998)\citenamefont{Martys, Shan, and
  Chen}}]{Martys98}
\bibinfo{author}{\bibfnamefont{N.~S.} \bibnamefont{Martys}},
  \bibinfo{author}{\bibfnamefont{X.}~\bibnamefont{Shan}}, \bibnamefont{and}
  \bibinfo{author}{\bibfnamefont{H.}~\bibnamefont{Chen}},
  \bibinfo{journal}{Phys. Rev. E} \textbf{\bibinfo{volume}{58}},
  \bibinfo{pages}{6855} (\bibinfo{year}{1998}).

\bibitem[{\citenamefont{Shan and Chen}(1993)}]{Shan93}
\bibinfo{author}{\bibfnamefont{X.}~\bibnamefont{Shan}} \bibnamefont{and}
  \bibinfo{author}{\bibfnamefont{H.}~\bibnamefont{Chen}},
  \bibinfo{journal}{Phys. Rev. E} \textbf{\bibinfo{volume}{47}},
  \bibinfo{pages}{1815} (\bibinfo{year}{1993}).

\bibitem[{\citenamefont{Ginzburg and Adler}(1994)}]{Ginzburg94}
\bibinfo{author}{\bibfnamefont{I.}~\bibnamefont{Ginzburg}} \bibnamefont{and}
  \bibinfo{author}{\bibfnamefont{P.~M.} \bibnamefont{Adler}},
  \bibinfo{journal}{J. Phys. II France} \textbf{\bibinfo{volume}{4}},
  \bibinfo{pages}{191} (\bibinfo{year}{1994}).

\bibitem[{\citenamefont{Ginzburg and d'{H}umi\`eres}(1995)}]{Ginzburg96}
\bibinfo{author}{\bibfnamefont{I.}~\bibnamefont{Ginzburg}} \bibnamefont{and}
  \bibinfo{author}{\bibfnamefont{D.}~\bibnamefont{d'{H}umi\`eres}},
  \bibinfo{journal}{J. Stat. Phys.} \textbf{\bibinfo{volume}{84}},
  \bibinfo{pages}{927} (\bibinfo{year}{1995}).

\bibitem[{\citenamefont{Chen et~al.}(1998)\citenamefont{Chen, Teixeira, and
  Molvig}}]{Chen98a}
\bibinfo{author}{\bibfnamefont{H.}~\bibnamefont{Chen}},
  \bibinfo{author}{\bibfnamefont{C.}~\bibnamefont{Teixeira}}, \bibnamefont{and}
  \bibinfo{author}{\bibfnamefont{K.}~\bibnamefont{Molvig}},
  \bibinfo{journal}{Intl. J. Mod. Phys. C} \textbf{\bibinfo{volume}{9}},
  \bibinfo{pages}{1281} (\bibinfo{year}{1998}).

\bibitem[{\citenamefont{Ansumali and Karlin}(2002)}]{Ansumali02a}
\bibinfo{author}{\bibfnamefont{S.}~\bibnamefont{Ansumali}} \bibnamefont{and}
  \bibinfo{author}{\bibfnamefont{I.~V.} \bibnamefont{Karlin}},
  \bibinfo{journal}{Phys. Rev. E} \textbf{\bibinfo{volume}{66}},
  \bibinfo{pages}{026311} (\bibinfo{year}{2002}).

\bibitem[{\citenamefont{Xu}(2003)}]{Xu03}
\bibinfo{author}{\bibfnamefont{K.}~\bibnamefont{Xu}}, \bibinfo{journal}{Phys.
  Fluids} \textbf{\bibinfo{volume}{15}}, \bibinfo{pages}{2077}
  (\bibinfo{year}{2003}).

\end{thebibliography}

\end{document}